\documentstyle[preprint,aps,epsfig]{revtex}
\tightenlines
\newcommand{\be}{\begin{equation}}
\newcommand{\ee}{\end{equation}}
\newcommand{\bea}{\begin{eqnarray}}
\newcommand{\eea}{\end{eqnarray}}

\def\simlt{\stackrel{<}{{}_\sim}}
\def\simgt{\stackrel{>}{{}_\sim}}
\begin{document}
\title{Electroweak baryogenesis induced by a scalar field}

\author{ Ram Brustein and David H. Oaknin }

\address{Department of Physics, Ben-Gurion University, 
Beer-Sheva 84105, Israel\\
email: ramyb,doaknin@bgumail.bgu.ac.il}

\maketitle
\begin{abstract}
A cosmological pseudoscalar field coupled to hypercharge topological  
number density can exponentially amplify hyperelectric and
hypermagnetic fields while coherently rolling or  oscillating, leading to the
formation of a time-dependent condensate of topological number density.
The topological condensate can be converted, under certain
conditions, into baryons in sufficient quantity to explain the observed
baryon asymmetry in the universe. The amplified hypermagnetic field can
perhaps sufficiently strengthen the electroweak phase transition,
and by doing so, save any  pre-existing baryon number asymmetry from
extinction. 
\end{abstract} 
\centerline{Preprint Number: BGU-PH-98/07}
\pacs{PACS numbers:  98.80.Cq,11.10.Wx,11.30.Fs,98.62.En}

To generate  baryon asymmetry from a state in which quark and antiquarks have
equal abundances,  the three
celebrated Sakharov conditions have to be
satisfied. In the scenario we propose, the Sakharov conditions are satisfied
during an epoch in which  a cosmological pseudo-scalar field coupled to
hypercharge topological number density coherently rolls or oscillates.  The scalar field
motion creates a time dependent hypercharge topological number condensate
which violates fermion number conservation through the abelian anomalous coupling
\cite{kuzmin}, and establishes the required departure from
equilibrium, and its interactions  violate C and CP symmetries. Pseudoscalar
fields with the proposed axion-like coupling appear
in several possible extensions of the Standard Model. They typically have
only perturbative derivative interactions and therefore vanishing potential at
high temperatures, and acquire a potential at lower temperatures through
non-perturbative interactions. Their potentials take the generic form
$V(\phi)=V_0^4 V(\phi/f)$, where $V$ is a bounded periodic function
characterized by a scale $f$ (``Peccei-Quinn" scale), which could be as high as
the Planck scale and a much smaller mass $m\sim V_0^2/f$, which could be as
low as a fraction of an eV, or as high as $10^{12}$ GeV. A particularly
interesting mass range is the TeV range, expected to appear if potential
generation is associated with supersymmetry breaking.
  
The fundamental role of hypermagnetic and hyperelectric fields in the  context
of electroweak (EW) baryogenesis has been recognized recently in several
investigations \cite{gs,enqvist}.It was observed that: {\em i)}   a topological
number condensate can be released at the EW transition in the form of leptons
and baryons, and  {\em ii)} strong enough
hypermagnetic fields could make the EW transition strongly first order even for
those large  values of the Higgs mass that have not been ruled out by LEP II
experiment.   Scalars with axion-like 
coupling to hypercharge fields were previously considered in \cite{widrow,guendelman}.  
Amplification of ordinary electromagnetic (EM) fields by such scalar fields was discussed in
\cite{widrow} and their possible use for baryogenesis in \cite{guendelman}.

We will assume that  the  universe is  homogeneous and isotropic, and can be
described by a conformally flat metric 
$ds^2 = a^2(\eta) (d\eta^2 - dx_1^2 - dx_2^2 - dx_3^2)$ where $a(\eta)$ is the
scale factor of the universe, and $\eta$ is conformal time related to cosmic
time $t$ as $a(\eta) d\eta= dt$.  In addition to the standard model fields we
will consider a  time-dependent pseudoscalar field $\phi(\eta)$ with coupling
$\frac{\lambda}{4} \phi Y\widetilde{Y}$ to the $U(1)_Y$ hypercharge field
strength and a
potential $V(\phi)=V_0^4 V(\phi/f)$. The coupling constant $\lambda$
(which we will take as positive)  has units of mass${}^{-1}$ $\lambda\sim
1/M$. For QCD
axions, $M\sim f$, but in general, it is not always the case, and we will
therefore take $M$ to be a free parameter, and  in particular
allow $M<f$. We will assume that the universe is  radiation
dominated at some early time $\eta=0$ at $T\simgt 100 GeV$ before the scalar
field becomes relevant.

Maxwell's equations describing the hyper EM fields (we will drop
the hyper from now on for brevity) in the resistive approximation \cite{biskamp} 
of the EW plasma,  coupled to the heavy pseudoscalar  are the following
 \begin{eqnarray}
&(i)&\ \nabla \cdot{\vec E} = 0 \hspace{.3in} 
(ii)\ \nabla \cdot {\vec B} = 0 \cr 
&(iii)&\ {\vec J} = \sigma {\vec E}  \hspace{.4in} 
(iv)\ \frac{\partial {\vec B}}{\partial \eta} = -\nabla \times {\vec E} \cr
&(v)&\ \frac{\partial {\vec E}}{\partial \eta} =  \nabla \times{\vec B} -
\lambda \frac{d \phi}{d \eta} {\vec B} - {\vec J}.\ \ \  
\label{maxwell}
\end{eqnarray}
We have rescaled the electric and magnetic fields 
$\vec E=a^2(\eta) \vec{\cal E}$, $\vec B=a^2(\eta) \vec{\cal B}$,  
 and the physical conductivity $\sigma=a(\eta)\sigma_c$. In the EW plasma
$\sigma_c \sim 10 T$ \cite{baym}. The fields $\vec{\cal E}$,  $\vec{\cal
B}$ are the flat
space EM fields.  We have assumed for simplicity vanishing  bulk velocity ${\vec
v}$ of the plasma and vanishing chemical potentials for all species. Our results
 can be easily generalized to  include non-zero bulk 
velocity or right electron chemical potential. The equation for the pseudoscalar
$\phi$ is the following   
\begin{equation} \label{axion} \frac{d^2 \phi}{d
\eta^2} + 2 a H \frac{d \phi}{d \eta} + a^2 \frac{d{\em V}(\phi)}{d\phi} =
\lambda a^2 {\vec E} \cdot{\vec B}, 
\end{equation}
where $H=\frac{1}{a^2} \frac{da}{d\eta}$ is the Hubble parameter. 
We will neglect the backreaction of the electromagnetic fields on the
scalar field since it is irrelevant for most of the physics we would like to
explore. The cosmic friction term 
$ 2 a H \frac{d \phi}{d \eta}$ will not be relevant for most of our
discussion.  We therefore solve eq.({\ref {axion}}) with vanishing r.h.s., and
substitute the resulting $\phi(\eta)$ into eq.({\ref{maxwell}). 

We are interested in solutions of the form $\vec E({\vec x},\eta)=
\int d^3{\vec k}\ e^{-i{\vec k}{\vec x}}\ {\vec e_{\vec k}}\
\epsilon_{\vec k}(\eta)$, $\vec B({\vec x},\eta)= \int d^3{\vec k}\
e^{-i{\vec k}{\vec x}}\ {\vec b_{\vec k}}\
\beta_{\vec k}(\eta)$, for which the electric and
magnetic modes are parallel to
each other.  We find that the Fourier modes ${\vec e}_{\vec k}$ and ${\vec
b}_{\vec k}$ are related, 
$
{{\vec b}_{\vec k}}{}^{\pm}= b_k^\pm  ({\hat e}_1 \pm i{\hat e}_2), 
\ \ \
k {{\vec e}_{\vec k}}{}^{\pm} \epsilon^\pm_{\vec k}(\eta)= 
\pm {{\vec b}_{\vec k}}{}^{\pm}  
\frac{\partial \beta_{\vec k}}{\partial\eta}^\pm,
$
where ${\hat e}_1$, ${\hat e}_2$ are unit vectors in the plane
perpendicular to ${\vec k}$ such that $({\hat e}_1,{\hat e}_2,{\hat k})$ is a
right-handed system. This type of solution has  $\vec J_{\vec k}\propto
\frac{\partial
\vec{B_{\vec k}}}{\partial\eta}$, $\nabla \times {\vec B_{\vec k}} \propto
\vec B_{\vec k}$, and $\nabla
\times {\vec E_{\vec k}} \propto \vec E_{\vec k}$. The function
$\beta_{\vec k}(\eta)$ obeys the following
equation,
\begin{equation}
\label{evolution}
\frac{\partial^2  \beta_k }{\partial \eta^2}^\pm+\sigma
\frac{\partial \beta_k}{\partial \eta}^\pm + 
\left( k^2 \pm \lambda \frac{d \phi}{d \eta} k \right)
\beta_k^\pm(\eta) = 0.
\end{equation}

Before plunging into a detailed discussion of the solutions of
eq.(\ref{evolution}), it is useful to consider the simple case of a constant
$\frac{d \phi}{d \eta}$. Then the solutions are simply
$
 \beta^\pm(\eta) = \beta_1^\pm\ \hbox{\it\large e}^{\hbox{$\omega_1^\pm\eta$}}
+  \beta_2^\pm\ \hbox{\it\large e}^{\hbox{$\omega_2^\pm\eta$}},
$
where $\omega_{1,2}^{\pm}$ are the two roots of the quadratic
equation
$
\label{roots}
\omega^2 + \sigma \omega + 
\left(k^2 \pm \lambda \frac{d \phi}{d \eta} k\right) =0 
$, 
$
\omega_{1,2}^\pm =
\frac{1}{2} \left[
-\sigma \pm \sqrt{\sigma^2 - 4\left(k^2\pm \lambda \frac{d \phi}{d \eta}
k\right) } \right].$

Now the qualitative behaviour of solutions is clear. If one of the eigenvalues
is positive, which can happen only if  
$\lambda\left| \frac{d \phi}{d \eta} \right|> k$, 
EM fields can grow exponentially. Otherwise fields are
either oscillating or damped, as in ordinary magnetohydrodynamics. To obtain
significant amplification, coherent scalar field velocities $\frac{d \phi}{d
\eta}$ over a  duration are necessary,  larger  velocities leading to larger 
amplification. We estimate the wavenumber $k$ of maximal amplification  by
looking for the maximum of $\omega$ as a function of $k$,  $k_{max}=\frac{1}{2}
\lambda \left|\frac{d \phi}{d \eta}\right|$. 

Another interesting approximate solution relevant for rolling fields can be
obtained for $k\ll T$. In this case, eq.(\ref{evolution}) can be approximated by
a first order equation  $\sigma\frac{\partial \beta}{\partial \eta}^\pm + 
\left( k^2 \pm \lambda \frac{d \phi}{d \eta} k \right) \beta^\pm(\eta)  = 0$,
which can be solved exactly, $\beta= \beta_0
\ \hbox{\it\large e}^{\hbox{$-\frac{k^2}{\sigma}\eta + \lambda \Delta \phi\
k/\sigma$}}$, where  $\Delta \phi(\eta)=|\phi(\eta)-\phi(0)|$. The
amplified mode is determined by the sign of  $\phi(\eta)-\phi(0)$. 
The amplification factor  
${{\cal A}^\pm}(k,\eta)=\beta^\pm(\eta)/\beta^\pm(0)$ is maximal
 for $k_{max}\eta=\frac{1}{2} \lambda \Delta \phi$,  
 ${{\cal A}^\pm}(k_{max},\eta)= e^{ \frac{1}{4} (\lambda \Delta \phi)^2 \frac{1}{\eta\sigma}}$.
Looking at  $\eta\sim \eta_{EW}$ we obtain  $\frac{1}{\eta_{EW}\sigma}\sim
10^{-16}$, and therefore to obtain amplification $\lambda \Delta
\phi\simgt 10^8$. For
specific models an upper bound on $\lambda \Delta \phi$ may appear, narrowing the range of
allowed parameter space. A value of $\lambda \Delta \phi\simgt 10^8$ is
not unnatural,
for example, such a value is obtained if the typical scale for scalar field
motion is the Planck scale, as happens in many models of supergravity, and
$\lambda\simlt 1/10^{10} GeV$.

\begin{figure}
\begin{center}
\psfig{figure=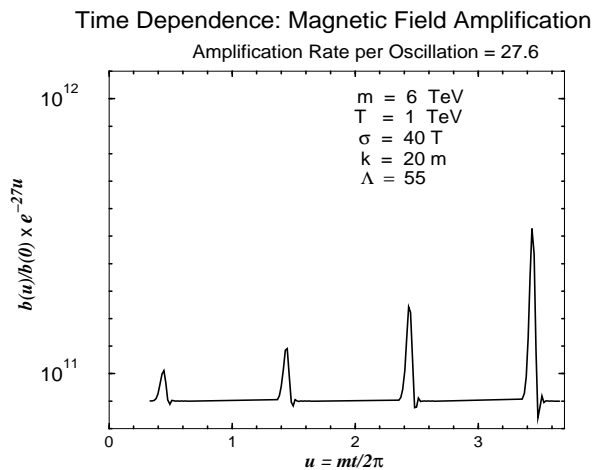,height=2.5in,width=3.2in}
\end{center}
\caption{Amplification of EM fields. The function $\beta$, exponentially
scaled,  is shown for the parameters $\Lambda=55$, $k/m=20$, $m=6 TeV$ and
$\sigma=40 T$ at $T=1 TeV$.
\label{fig:amp} }
\end{figure}

If the scalar field is oscillating, the field's
velocity changes sign periodically and both modes can be amplified, each during a
different part of the cycle.  Each mode is amplified during one part of the
cycle and damped during the other part of the cycle. Net amplification results
when amplification overcomes damping, which occurs roughly when 
$\lambda \frac{d \phi}{d \eta} k \simgt \sigma^2$ (recall that $\sigma\sim
10T$). Total amplification is exponential in the number of cycles. 

We have studied numerically the solutions of eq.(\ref{evolution}) for the
case of an oscillating field for different parameters and situations. The
concrete potential that we used for the numerical analysis is $V(\phi)=\frac{1}{2} m^2
\phi^2$, assuming that the initial amplitude of the field is of order $f$ with
$\Lambda\equiv f/M>1$, $f/m>1$. We find results in agreement with the previous
qualitative discussion. Amplification occurs for a limited range of Fourier
modes, peaked around  $k/m \sim \Lambda$. The modes of the EM fields are
oscillating with (sometimes complicated) periodic time dependence and an
exponentially growing amplitude. In Fig. \ref{fig:amp}  we show an example of
the time dependence of amplified EM fields for a specific
mode. For the range of parameters in which fields are amplified, the amount of
amplification per cycle for each of the two modes is
very well approximated by
the same  constant $\Gamma(k/m,\Lambda,\sigma)$. A good approximate estimate for
the average amplification after $N$ cycles  is therefore 
${{\cal A}^\pm}(k,\eta)={{\cal N}^\pm}_{k} e^{N \Gamma}$, where
${{\cal N}^\pm}_{k}$ represents the transient influence of the initial
conditions of EM and scalar fields.  

The scalar field is a
very efficient amplifier of EM fields. For example, to obtain an amplification
of $10^{12}$ for $\Lambda=55$, $k/m=20$, $m=6 TeV$ and $\sigma=40 T$, for
oscillations occurring at a temperature of $1 TeV$, we need just one cycle!
 In Fig. \ref{fig:amppercycle} we have presented an example of the 
dependence of the amplification per cycle factor $\Gamma$, on different
parameters.

\begin{figure}
\begin{center}
\psfig{figure=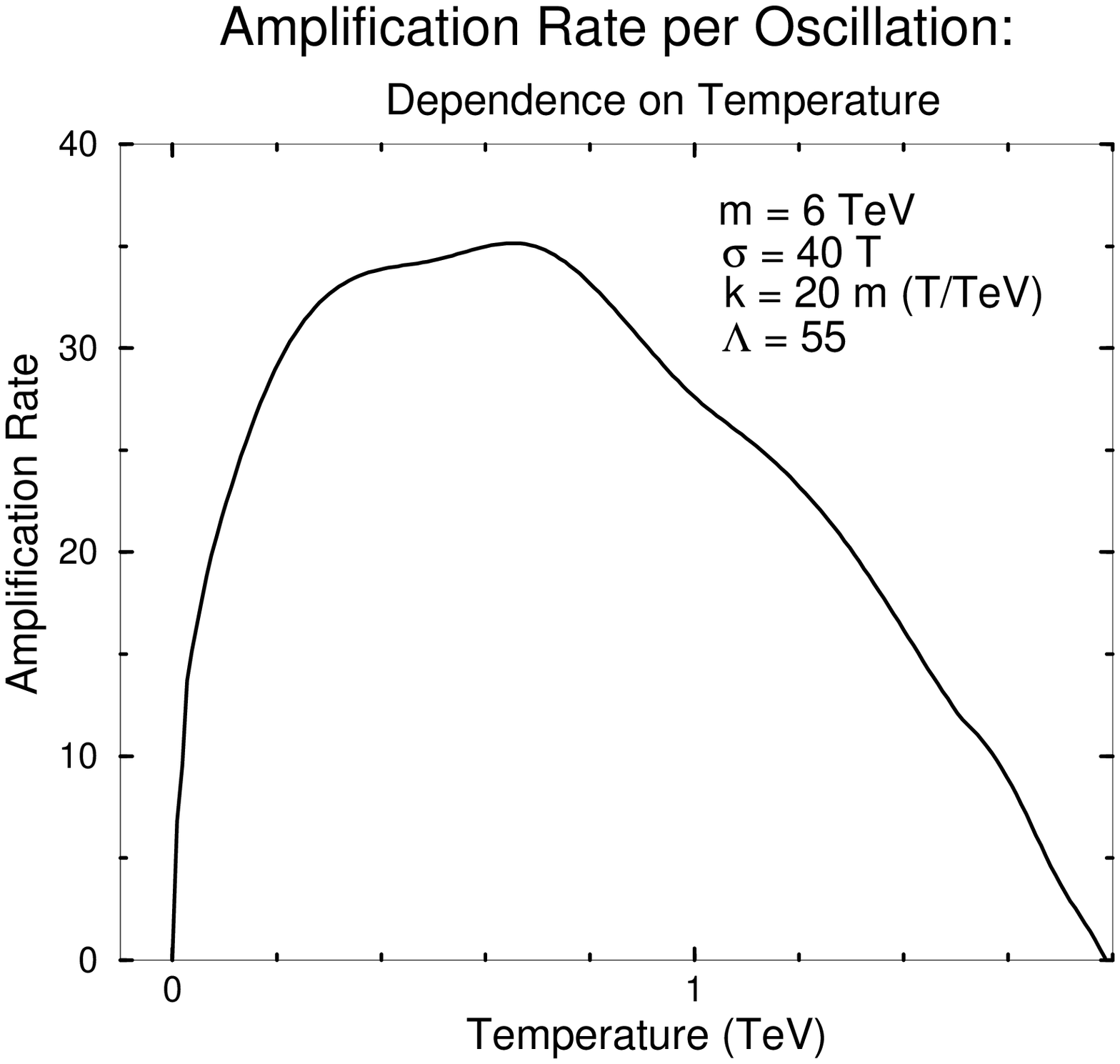,height=2.5in,width=3.2in} 
\psfig{figure=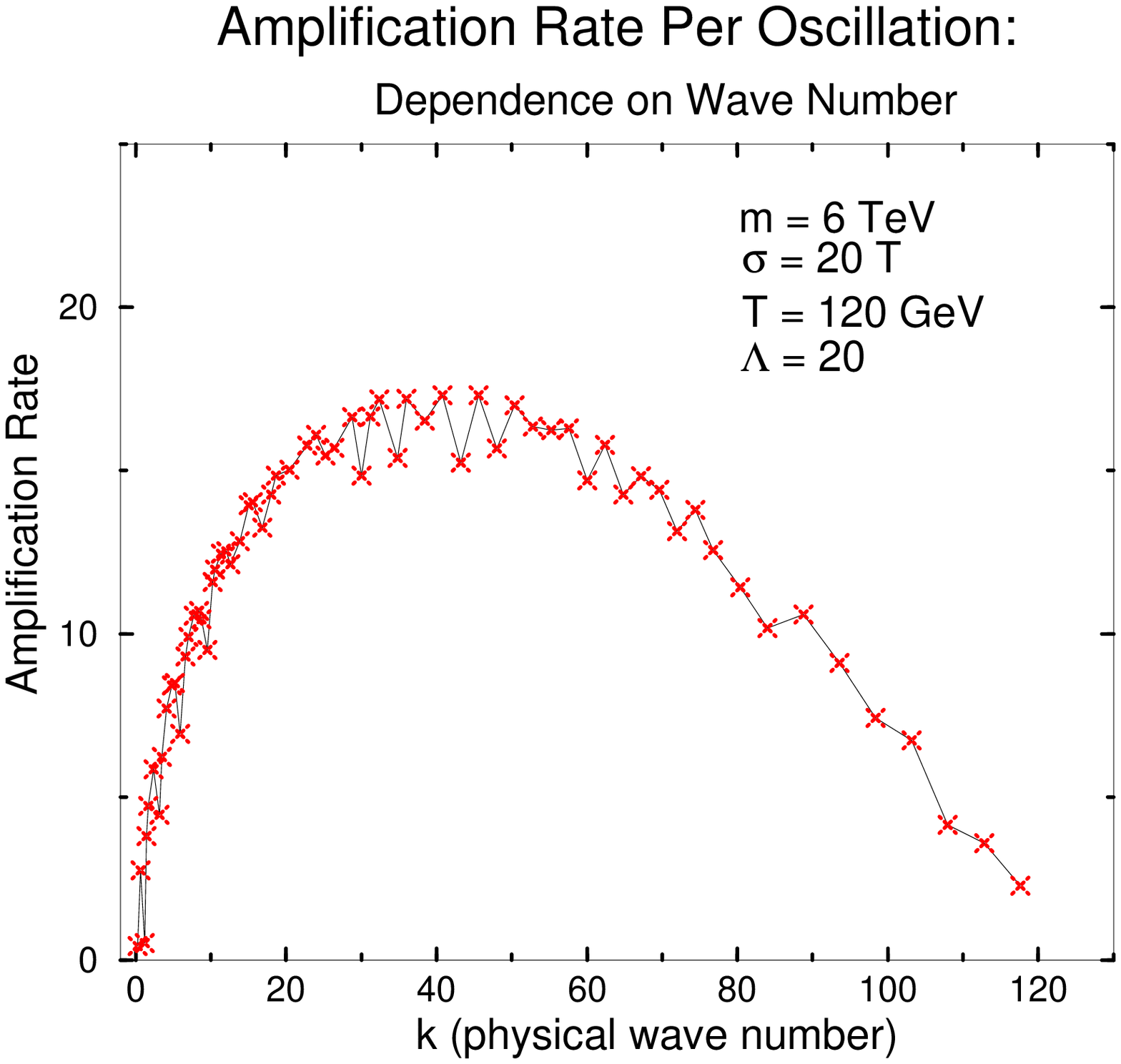,height=2.5in,width=3.2in} 
\psfig{figure=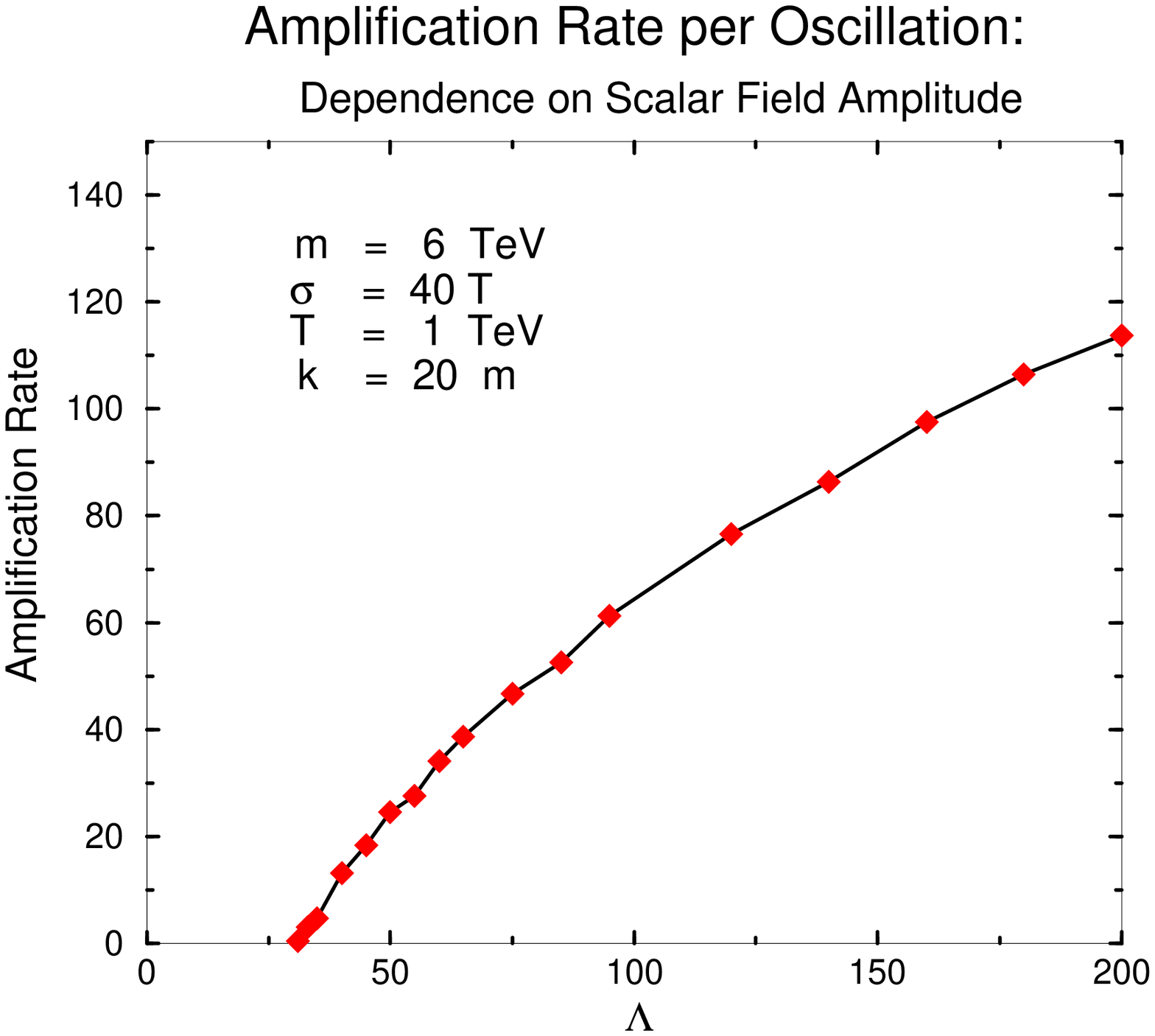,height=2.5in,width=3.2in}
\end{center}
\caption{Amplification per cycle of EM fields, as a function of temperature,
wave number, and scalar field amplitude.
 \label{fig:amppercycle} }
\end{figure}

A detailed discussion of the end of oscillations or rolling
is beyond the scope of the current investigation. However, we do know
 that once
the oscillations or rolling stop, the fields are no longer amplified and
obey a diffusion equation \cite{gs}. Modes with wave number below the diffusion value
$k <k_\sigma T\sim 10^{-8} T$, where
$\frac{k_{\sigma}^2}{\sigma}\frac{1}{\eta_{EW}}=1$, remain almost
constant until the EW transition, their
amplitude goes down as $T^2$, and energy density as $T^4$, maintaining a
constant ratio with the environment radiation. Modes with $k/T>k_\sigma$
decay
quickly, washing out the results of amplification.
We have seen that the range of amplified momenta for oscillating fields is not
too different than $T$, therefore scalar field  oscillations have to occur just
before, or during the EW transition. In that case, the amplified fields do not
have enough time to be damped by diffusion. If the field is rolling, momenta
$k\ll T$ can be amplified, and therefore the rolling can end sometime before the
EW transition.

To obtain the average magnetic energy density in amplified fields we have to
average over the magnetic and scalar fields initial conditions. 
The averaging has different reasons: {\em i)} statistical
fluctuations inside the horizon of magnetic field initial conditions
due to thermal and quantum fluctuations,  and {\em ii)} variations over
causally disconnected regions. We will assume translation and rotation
invariance of initial conditions, 
$
\langle B^{\pm}({\vec k}, \eta) {B^{\pm}({\vec \ell}, \eta)}^* \rangle_{|\eta=0}=
\delta^3({\vec k}-{\vec \ell})\ \delta_{+ -} f^\pm(k).
$
Using the magnetic power spectrum
$
P_B^\pm(k)=4\pi  k^3 f^\pm(k),
$
the average initial energy density $\rho_B(0)$, given by
$\rho_B(0) =\frac{1}{8 \pi} <B^2>_{|\eta=0}$, can be expressed as
$
\rho_B(0)=\frac{1}{8\pi}\int\limits_0^\infty d \ln k  \left( P_B^+(k) +
P_B^-(k)\right).
$

The two modes of the magnetic field are amplified by amplification factors  
${{\cal A}^\pm}({k},\eta)$,   which  depend
on the parameters of the model as discussed previously.
The amplification factors depend, in addition, on the initial conditions of the
magnetic field and on the initial conditions of the scalar field. For example,
if the velocity of the scalar field $d\phi/d\eta$ is constant, the sign of the
velocity determines that only ${\cal A}^+$ or ${\cal A}^-$ will be
non-vanishing.  

The magnetic energy density in amplified fields is given by
\be
\rho_B= \frac{1}{8 \pi}  \int d\ln k \Biggl\{ 
P_B^-(k) \left| {{\cal A}^-}({k},\eta)\right|^2 +
 P_B^+(k)  
 \left| {{\cal A}^+}({k},\eta)\right|^2
\Biggr\}.
\ee
The Chern-Simons number density is given by
$
\Delta n_{CS}=- \frac{y_R^2 g'^2}{16 \pi^2}  \int\limits_0^\eta 
d{\widetilde \eta} \langle E\cdot B\rangle .
$, where $y_R=-2$ is the hypercharge of the right electron, 
and $g'$ is the hypercharge gauge coupling.
Since in our case $E_{ {\vec k}}^\pm(\eta)=
\pm \frac{1}{k} \partial_\eta B_{ {\vec k}}^\pm(\eta)$, then 
$
\langle E^{\pm}({\vec k}, \eta) {B^{\pm}({\vec \ell}, \eta)}^* \rangle=
\pm \frac{1}{2} \frac{1}{k} \partial_\eta \langle B^{\pm}({\vec k}, \eta)
{B^{\pm}({\vec
\ell}, \eta)}^* \rangle.
 $
Neglecting the initial $n_{CS}$ density, 
\be
 n_{CS}= \frac{y_R^2 g'^2}{32 \pi^2}   
\int d\ln k \frac{1}{k}  \Biggl\{ 
P_B^-(k) \left| {{\cal A}^-}({{\vec k}},\eta)\right|^2 -
 P_B^+(k)  
 \left| {{\cal A}^+}({{\vec k}},\eta)\right|^2
\Biggr\}.
\ee

Let us define two k-dependent asymmetry parameters,
$
{ \gamma}^B_{AS}= \frac{P_B^-(k) - P_B^+(k)} {P_B^-(k) + P_B^+(k)},$ and 
$\gamma^\phi_{AS} = \frac{\left| {{\cal A}^-}({k},\eta) \right|^2 -
\left| {{\cal A}^+}({k},\eta) \right|^2} 
{\left| {{\cal A}^-}({k},\eta)  \right|^2 +
\left| {{\cal A}^+}({k},\eta) \right|^2}.
$
Both parameters vary from 0 - no asymmetry, to $\pm 1$ - maximum
asymmetry. 
Using the asymmetry parameters we may relate $n_{CS}$ to $\rho_B$,
\be
n_{CS}= \frac{y_R^2 g'^2}{4 \pi}  \int d \ln k \frac{1}{k} \rho_B (k) 
\frac{\gamma^B_{AS}+ \gamma^\phi_{AS} } {1+ \gamma^B_{AS} \gamma^\phi_{AS} }.
\ee
Further, we can compute the fractional energy density in coherent magnetic field
configurations  $\Omega_B(k)=\rho_B(k)/\rho_c$, where $\rho_c$ is the critical
energy density, and the Chern-Simons fractional number density $n_{CS}/s$,
where $s$ is the entropy density, and compare them. If the universe is radiation
dominated $\rho_c=\frac{\pi^2}{30} g_* T^4$ and $s=\frac{2\pi^2}{45}g_{*s} T^3$,
therefore   
\be
 \frac{n_{CS}}{s}= \frac{y_R^2
g'^2}{4 \pi} \frac{3 g_*}{ 4 g_{*s}}  \int d \ln k \frac{T}{k} \Omega_B
(k) 
\frac{\gamma^B_{AS}+ \gamma^\phi_{AS} } {1+ \gamma^B_{AS} \gamma^\phi_{AS} }.  
\ee
If the amplification factors ${{\cal A}^\pm}_{k}$ are, as we have seen,
sharp
functions of $k$, peaked at $k_{max}$ and have width in $k$ of approximately
$k_{max}$, then 
\be
\frac{n_{CS}}{s}\simeq 0.01 
 \frac{T}{k_{max}} \Omega_B(k_{max})  \frac{\gamma^B_{AS}+
\gamma^\phi_{AS} } {1+ \gamma^B_{AS} \gamma^\phi_{AS} } (k_{max}),
\ee
where we have used numerical values for the prefactor coefficients.

According to Giovannini and Shaposhnikov \cite{gs}, this Chern-Simons number
will be released in the form of fermions which will not be erased if the EW
transition is strongly first order\cite{shaposhnikov}, and will
generate a baryon
asymmetry, 
$\frac{n_{B}}{s}=-\frac{3}{2} \frac{n_{CS}}{s}$.  An equal lepton number would
also be generated by the same mechanism so that $B-L$ is conserved.
Note that the fact that baryon number asymmetry is generated at $k_{max}\ne 0$
does not mean that baryon density is actually inhomogeneous on this
short length scale $L_{max} \sim 1/k_{max}$. Comoving neutron diffusion
distance at the beginning of nucleosynthesis is much longer than
$L_{max}$ \cite{jedamzik,applegate}, so that by that time inhomogeneities
would have been erased by free streaming \cite{jedamzik}. 

If $T/ k_{max}$ is not too different than unity, as we have seen for the case of
oscillating field, and $\gamma^B_{AS}$ and  $\gamma^\phi_{AS}$ are small, 
it is possible to obtain $\frac{n_{B}}{s}\sim 10^{-10}$ and have strong magnetic 
fields $\Omega_B\sim 1$ present during the EW transition. If $T/ k_{max}$
is large and $\gamma^\phi_{AS}$ is order unity as
we have seen in the rolling case,  it is not possible to have 
strong magnetic fields without producing too many baryons.

The existence of hypermagnetic fields during the EW transition can influence the
nature of the EW transition. For example, a homogeneous hypermagnetic field
adds a pressure term to the symmetric phase which could lower the
transition temperature and strengthen the EW transition. This well known
effect in conductor-superconductor phase transitions
\cite{meissner} was
discussed in several investigations \cite{gs,enqvist}, with seemingly
inconclusive results, for the moment.  

Our conclusions are therefore that\\
1. Depending on the ratio $T/ k_{max}$ and on the asymmetry
parameters, it is possible to create the desired ratio $\frac{n_{CS}}{s} \sim
10^{-10}$, and therefore to generate the observed baryon asymmetry in the
universe. \\
2. Some asymmetry in the initial conditions of either
$B$ or $\phi$ is required. The asymmetry can be very small and can be
induced at a much higher scale. Possible 
sources for the initial asymmetry are temperature dependent potential that traps
$\phi$ at a preferred position, asymmetry in quantum fluctuations, etc. Large
asymmetry  in $\phi$ can be expected and appears naturally.

\acknowledgments 
This work is supported in part by the  Israel
Science Foundation administered by the Israel Academy of Sciences and
Humanities. D.O. is supported in part by the Ministry of Education and
Science of Spain.


\begin{references}

\bibitem{kuzmin} V.A. Kuzmin, V.A. Rubakov and M.E. Shaposhnikov, {\it
Phys. Lett.} {\bf B155} (1985) 36.

\bibitem{gs} M. Giovannini and M.E. Shaposhnikov, {\it Phys. Rev.} {\bf
D57} (1998) 2186; M. Giovannini and M.E. Shaposhnikov, {\it Phys.
Rev. Lett.} {\bf 80} (1998) 22; M. Joyce and M.E. Shaposhnikov, {\it
Phys. Rev. Lett.} {\bf 79} (1997) 1193.

\bibitem{enqvist} P. Elmfors, K. Enqvist and K. Kainulainen,
hep-ph/9806403.

\bibitem{widrow} M.S. Turner and L.M. Widrow, {\it Phys. Rev.} {\bf
D37} (1988) 2743; W.D. Garretson, G.B. Field and S.M. Carroll, {\it Phys.
Rev.} {\bf D46} (1992) 5346.

\bibitem{guendelman} E.I. Guendelman and D.A.Owen, {\it Phys. Lett.} {\bf
B276} (1992) 108.


\bibitem {biskamp} D. Biskamp, {\it Nonlinear Magnetohydrodynamics},
Cambridge University Press, Cambridge, 1994.

\bibitem{baym} G. Baym and H. Heiselberg, {\it Phys. Rev.} {\bf D56} (1997) 5254; 
M. Joyce, T. Prokopec and N. Turok, {\it Phys. Rev.} {\bf D53} (1996) 2930.



\bibitem{shaposhnikov} M.E. Shaposhnikov, JETP {\it Lett.} {\bf 44} (1986) 465; 
{\it Nucl. Phys.} {\bf B287} (1987) 757.

\bibitem{jedamzik} K. Jedamzik and G.M. Fuller, {\it Astrophys. J.} {\bf
423} (1994) 33; K. Jedamzik, G.M. Fuller and G.J. Mathews, {\it Astrophys.
J.} {\bf 43} (1994) 50; K. Kainulainen, H. Kurki-Suonio and E. Sihvola,
astro-ph/9807098.

\bibitem{applegate} H. Kurki-Suonio, K. Jedamzik and G.J. Mathews, {\it
Astrophys. J.} {\bf 479} (1997) 31; J. Applegate, C.H. Hogan and R.J.
Scherrer, {\it Phys. Rev.} {\bf D35} (1987) 1151.

\bibitem{meissner} E. Lifshitz and L.P. Pitaevskii, {\it Statistical
Physics, Part 2}, Pergamonn Press, Oxford, 1981; A.L. Fetter and J.D.
Walecka, {\it Quantum Theory of Many Particle Systems}, McGraw-Hill,
New-York, 1971.


\end{references}
\end{document}